\documentclass[onecolumn,12pt]{IEEEtran}

\usepackage{algorithm,algorithmic,amsbsy,amsmath,amssymb,epsfig}

\DeclareMathAlphabet{\mathpzc}{OT1}{pzc}{m}{it}
\usepackage{times,amsmath,epsfig}
\usepackage{cite}
\usepackage{graphicx}
\usepackage{psfrag}
\usepackage{subfigure}
\usepackage{url}
\usepackage{stfloats}
\usepackage{amsmath}
\usepackage{amssymb}
\usepackage{array}
\usepackage{array}

\newcommand{\beq}{\begin{equation}}
\newcommand{\eeq}{\end{equation}}
\newcommand{\bitm}{\begin{itemize}}
\newcommand{\ba}{\begin{array}}
\newcommand{\ea}{\end{array}}
\newcommand{\eitm}{\end{itemize}}
\newcommand{\beqn}{\begin{eqnarray}}
\newcommand{\eeqn}{\end{eqnarray}}
\newcommand{\beqno}{\begin{eqnarray*}}
\newcommand{\eeqno}{\end{eqnarray*}}
\newcommand{\bma}{\begin{displaymath}}
\newcommand{\ema}{\end{displaymath}}
\newcommand{\bnu}{\begin{enumerate}}
\newcommand{\enu}{\end{enumerate}}
\newcommand{\bce}{\begin{center}}
\newcommand{\ece}{\end{center}}
\newcommand{\btb}{\begin{tabular}}
\newcommand{\etb}{\end{tabular}}

\begin{document}
%
% paper title

\title{Resource Management in Non-orthogonal Multiple Access
Networks for 5G and Beyond}
%\title{Non-orthogonal Multiple Access: Taxonomy and Design}
\author{
\IEEEauthorblockN{Lingyang Song\IEEEauthorrefmark{1}, Yonghui Li\IEEEauthorrefmark{2}, Zhiguo Ding\IEEEauthorrefmark{3} and H. Vincent Poor\IEEEauthorrefmark{4}\\}
\IEEEauthorblockA{\IEEEauthorrefmark{1}School of Electrical Engineering and Computer Science, Peking University, China\\}
\IEEEauthorblockA{\IEEEauthorrefmark{2}School of Electrical and Information Engineering, The University of Sydney, Australia\\}
\IEEEauthorblockA{\IEEEauthorrefmark{3}School of Computing and Communications, Lancaster University, UK\\}
\IEEEauthorblockA{\IEEEauthorrefmark{4}Department of Electrical Engineering, Princeton University, Princeton, US}
}

% make the title area
\maketitle

\thispagestyle{empty}
\begin{abstract}

Non-orthogonal multiple access (NOMA) techniques have been recently proposed for the small-cell networks in 5G to improve access efficiency by allowing many users to share the same spectrum in a non-orthogonal way. Due to the strong co-channel interference among mobile users introduced by NOMA, it poses significant challenges for system design and resource management. This article reviews resource management issues in NOMA systems. The main taxonomy of NOMA is presented by focusing on the following categories of resources reuse:  power-domain and code-domain NOMA. Then a novel radio resource management framework is presented based on game-theoretic models for uplink and downlink transmissions. Finally, potential applications and open research directions in the area of resource management for NOMA are provided.
\end{abstract}

\newpage \setcounter{page}{1}\setlength{\baselineskip}{25pt}

%%%%%%%%%%%%%%%%%%%%%%%%%%%%%%%%%%%%%%%%%%%%%%%%%%%%%%%%%%%%%%%%%%%%%%%%%%%%%%%%%%%%%%%%%%%%%%%%%%%
\section{Introduction}
%%%%%%%%%%%%%%%%%%%%%%%%%%%%%%%%%%%%%%%%%%%%%%%%%%%%%%%%%%%%%%%%%%%%%%%%%%%%%%%%%%%%%%%%%%%%%%%%%%%

In existing wireless systems, such as fourth-generation (4G) long term evolution (LTE), orthogonal frequency division multiplexing access~(OFDMA) has been widely utilized to schedule different users' transmissions in orthogonal temporal and frequency domains~\cite{Song2010}. Due to an explosively increasing demand for wireless traffic, future fifth generation (5G) wireless systems will face greater challenges and demand a much higher spectral efficiency, massive connectivity, and low latency. However, in conventional  orthogonal multiple access~(OMA) systems,  mobile devices share resources in an orthogonal manner, and these OMA schemes will suffer from serious congestion problems when there are a large number of access devices due to the limited transmission bandwidth. Thus,  much more spectrally efficient radio access technologies are required for future mobile communication systems.

To tackle these demands for massive connectivity and broadband services, non-orthogonal multiple access (NOMA) techniques have recently been introduced at the small-cell networks, which allow users to access the channel non-orthogonally by either power-domain multiplexing~\cite{YS2012,DZ2014} or code-domain multiplexing~\cite{JB2009,HN2013}. Unlike conventional  OMA systems in which  one subcarrier is  assigned to only one user, NOMA allows multiple number of users to share the same subcarriers, which introduces co-user interference. To cope with this problem, multi-user detection (MUD) techniques to reduce the co-channel interference, such as successive interference cancelation (SIC), need to be applied at the receiver side to decode the received signals at the cost of increased computational complexity.

The superior spectral efficiency of NOMA lies in the fact that NOMA exploits  a smart reuse of the network resources, by multiplexing  multiple users' messages in the same sub-channel at the transmitter, and constructing effective signal detection algorithms at the receivers. As such, NOMA requires a new design of physical-layer transmission, MAC and network resource allocation algorithms. This promising opportunity has been clearly  demonstrated by  recent  research developments in this area. For example, in~\cite{YYATAK-2013}, the authors discussed how to realize  user fairness in uplink wireless networks. A low-complexity power control based on a tree search algorithm with a successive interference cancellation~(SIC) receiver was discussed in~\cite{AAH-2014}. Though there are some initial studies on resource management in NOMA systems ~\cite{YYATAK-2013}\cite{AAH-2014}, they have mainly focused on  power allocation for power-domain NOMA, whereas other types of network resources in both frequency and code dimensions can be further explored in NOMA networks to increase the access network capacity.

In this article, we will discuss various technical issues in the NOMA access protocol design and resource allocation algorithms. In particular, we mainly discuss the following two categories of NOMA:
\begin{itemize}
    \item {\em Power-domain~(PD) NOMA}: PD-NOMA multiplex multiple users in the same subcarrier simultaneously with different power level at the transmitter, and SIC can be applied at the receiver side to decode the received signals~\cite{YS2012}.
    \item {\em Code-domain~(CD) NOMA}: CD-NOMA uses  sparse (or low-correlation) spreading sequences to spread data of each user over multiple subcarriers to realize overloading~\cite{JB2009}. It can be also transformed to codebook based schemes, in which the codewords can be generated by those spreading sequences known by both the transmitter and receivers~\cite{HN2013} .
\end{itemize}
Obviously, in different NOMA protocols, the network resources such as power, frequency, and time, need to be optimized by exploring proper optimization algorithms. In this article, we present system design   and resource allocation issues  for both PD and CD NOMA protocols, and discuss  recent promising research developments in this rapidly growing research area.

The rest of this article is organized as follows: Section~II presents various transceiver design issues in NOMA. The radio resource allocation problems for PD- and CD-NOMA are presented in Section III and IV, respectively. In Section~V, we   draw  some conclusions, and also discuss potential applications and research directions in NOMA communications.

\section{NOMA Transceiver Design in Power and Code Domains}

NOMA increases the access spectral efficiency of uplink and downlink transmission in wireless networks by exploiting the new multiplexing protocols in various domains such as power and frequency. As a new multiple access protocol beyond the conventional OMA mechanism, NOMA requires a new design of access protocols, transceiver optimization and resource allocation. In this section, we will discuss the design issues for PD-NOMA and CD-NOMA, respectively.

\subsection{Transceiver Design for Power-domain NOMA}

In a PD-NOMA cellular system with one base station~(BS) and a set of users, multiple users are multiplexed at the same subcarrier at the same time. Specifically, in the downlink transmission, the BS broadcasts a superposition of multiple users' signals by properly choosing  the transmit power coefficients for these users' signals subject to a total power constraint. In general, lower power levels will be allocated to the users with better channel gains while larger power assignment levels will be allocated to the users with worse  channel gains. Since a subcarrier can be occupied  by multiple number of users, the signal of one user will cause  interference to the others. In order for a user to recover its signal from the superimposed signals, SIC is applied to the received signals. For SIC, the signals of users with channel gains lower than that of the desired user will be recovered sequentially  in the order of the channel gains, and that user treats the signals of users with channel gains higher  than its channel gain as noise. In the uplink multiple access channels, multiple mobile users transmit their own signals to the BS. The multiple users' signals are then recovered at the BS by using SIC. The sum-rate performance of NOMA depends on the power control policy used at the users. Based on the feedback from the BS, iterative distributed power control at the user side needs to be performed subject to individual power constraints.

From an information theoretic perspective, PD-NOMA with a SIC receiver can approach the multiuser capacity region in the downlink broadcast channels. This becomes a particularly important application for cellular systems where the channel quality vary significantly between users in the cell center and those at the cell edge due to the near-far effect. Different from conventional power adjustment strategies, such as water filling for MIMO, in NOMA, users with worse channel gains obtain more transmit power. Specifically, the signals to a user with the poor channel conditions are allocated with more transmit power, which ensures that that user can detect its message directly by treating other users' signals as noise. On the other hand, a user with  a stronger channel condition needs to first decode information of other users with weaker channels and then subtract them from the received signals before decoding its own message, using SIC.

\subsection{Transceiver Design for Code-domain NOMA}

CD-NOMA can be treated as an open-loop scheme to accommodate multiple users over shared time and frequency resources, which spreads data streams of each user over multiple subcarrier by sparse or low-correlation spreading sequences. Thus, spreading signals over OFDM tones can greatly improve the quality of transmission procedure due to the interference averaging and whitening as well as additional frequency diversity. At the receiver, multi-user detection algorithms such as the near-optimal message passing algorithm (MPA) can be adopted to recover the original signals. To facilitate the control, predefined codebook methods can be applied~\cite{HN2013}, in which the codewords are generated through spreading symbols with different power levels such as QAM or specifically designed constellation using low-density spreading sequences with a few nonzero elements within a large signature length. Selected from the codebook sets, the incoming bits can be then directly mapped to multi-dimensional complex codewords. Note that in the downlink scenario, the codeword selection and power level adjustment can be performed in a centralized way.

For uplink CD-NOMA, contention based transmission schemes can be considered to decrease the latency and signaling overhead. The mobile users send data in such a way that they can contend for multiple physical resources such as time and frequency without relying on the centralized request and grant procedure. Specifically, the assigned time and frequency resources are shared by a number of users that can use a specific set of resources in contention transmission. When a user has data to send, it attempts to deliver the packet through the radio resource reserved for contention transmission. This is particularly important for mission critical applications with ultra-low latency requirements. Similar to CDMA, CD-NOMA can be considered as a  type of repetition coding in which transmitted symbols are modulated by spreading sequences. However, unlike conventional CDMA systems, CD-NOMA transmits signals in the form of OFDM symbols across the subcarrier by discrete Fourier transformation in order to be compatible with current LTE systems, and thus, power allocation or control at the subcarrier level is still required.

\section{Game-theoretic Resource Allocation Models for Power-Domain NOMA}

This section discusses game-theoretic models used to solve resource allocation problems in uplink and downlink PD-NOMA systems.

\subsection{Noncooperative Power Control Game for Uplink PD-NOMA}

Power control techniques have been widely used in multi-user mobile communication systems to minimize the multi-user interference and optimize the link data rate. Power control is particularly important for uplink PD-NOMA, because the transmit power of the mobile users needs to be properly adjusted in order to optimize the interference among co-channel users. Since users are distributed in the network and coordinated power control will entail too much overhead, interactions and power control among the PD-NOMA users sharing the same channels can be modeled as a noncooperative game implemented in a distributed way. Specifically, the players in the game are the mobile users, the action is to control the transmit power in each channel subject to the power constraint, and the payoff function is defined as the rate minus the cost of the power used by the mobile user transmission. Thus, a distributed power control algorithm can be designed by iteratively updating the transmit power of each user given the individual power constraint. The major difference compared to traditional distributed power control lies in that each user may occupy a different number of subcarrier and may not align with each other in the frequency domain, such that one user may collide partially in the frequency band with others and result in different levels of interference in each subcarrier. Hence, from the per subcarrier point of view, the above power control game can be applied, but we still need to consider the total power allocation over all the occupied subcarriers.

\subsection{Coalition Formation Game Model for Downlink PD-NOMA}

\subsubsection{Coalition Game Basics}
In a coalitional game, a set of players that form cooperative groups are called coalitions \cite{Han2011}. The reason for these users to join in the same coalition is that these users can gain higher payoffs, and typically the sum of the these payoffs can be defined as the coalitional value. Coalition formation games can be classified into two kinds: the strategic form and partition form games. In the strategic form, the coalition value only relies on the payoffs of the members in the same coalition, while in the partition form, the value of a coalition becomes quite involved and  depends on the actions of the players outside the coalition. Coalitional games have attracted a lot of attention in the research community in wireless communications, since they are the ideal tools to realize user clustering in a large scale wireless network.

%For the strategic-form coalitional formation games, a typical algorithm is the merge-and-split algorithm with the  two following operations:
%\begin{itemize}
% \item \emph{Merge:} Coalitions merge to a single coalition whenever mutual benefits exist.
% \item \emph{Split:} A coalition splits whenever this splitting can provide better payoffs.
%\end{itemize}
%
%It has been proved that the merge-and-split algorithm will always converge to a set of stable coalitions, in which no individual player has interests in changing coalition through a merge or split operation to achieve a higher utility.

\subsubsection{Coalition Game for Downlink PD-NOMA User Grouping}

\begin{figure}[!t]
\centering
\includegraphics[width=4.5in]{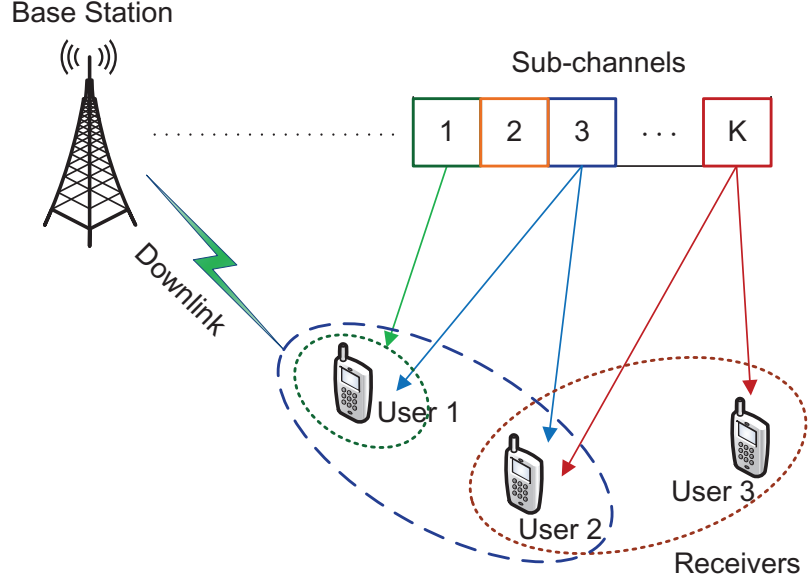}
\caption{System model of a PD-NOMA network.} \label{system_model_pdnoma}
\end{figure}

In this section, we discuss how the coalition game can be used to solve the downlink PD-NOMA user grouping. Consider a downlink NOMA network with only one cell and a number of mobile users, given in Fig.~\ref{system_model_pdnoma}, in which the BS transmits the signals to serve the mobile users denoted by ${\cal{N}} = \left\{ {1, \cdots ,N} \right\}$. The BS divides the available bandwidth into a set of subcarriers, denoted by ${\cal{K}} = \left\{ {1, \cdots ,K} \right\}$, where $N$ is much larger than $K$. Notice that other types of network resources, such as multiple antennas, time slots and spreading codes, can be also directly applied to avoid inter-group interference.  When the BS has the full knowledge of the channel side information (CSI), the BS assigns a subset of non-overlapping subcarriers to the users, and allocates different levels of power to the users subject to the total power constraint. According to the before mentioned PD-NOMA protocol, one subcarrier can be scheduled with multiple users, and meanwhile, one user can obtain the same transmitted signals from the BS through multiple subcarriers. This can be formulated as a coalition game.

The design of game theoretic approaches for the aforementioned scenario is challenging since downlink PD-NOMA user grouping is likely to be in a partition form for which many well-known approaches, such as merge-and-split algorithms, cannot be applied. For example, consider a scenario which has $|\mathcal{K}|$ coalitions, denoted by $\mathcal{S}_k$, $k\in\{1, \cdots, |\mathcal{K}|\}$, where there is only one user in $\mathcal{S}_1$. In the case that the user in $\mathcal{S}_1$ decides to join in   coalition $\mathcal{S}_2$, only $(|\mathcal{K}|-1)$ subcarriers are needed, instead of $|\mathcal{K}|$. Note that in the considered NOMA system,  user $i$ in coalition $k$ has a rate of $\frac{1}{\mathcal{N}}\log(1+\text{SINR}_{i,k})$, where $\mathcal{N}$ denotes the total number of the used subcarriers, and $\text{SINR}_{i,k}$ denotes the signal-to-interference-plus-noise ratio (SINR) of this user. Therefore, the action that the user in $\mathcal{S}_1$ moves to $\mathcal{S}_2$ has an impact on the rates not only for the users in $\mathcal{S}_1$ and $\mathcal{S}_2$, but also for those in other coalitions, which means that the modelled game is in a  partition form.

A promising solution for the aforementioned challenge is to use the fact that users in many communication scenarios, such as 5G systems and the Internet of Things (IoT), have dynamic quality of service (QoS) requirements. In particular, consider a scenario in which there are two categories of users. The first category of users need to be served for small packet transmission in a quick manner. Examples for such users can be health-care sensors which need to be connected for safety-critical applications, and connections with low data rates are sufficient to them. The second category of users can be broadband users who are performing background tasks, such as web browsing and movie downloading. The use of OMA results in the situation that the sensors get more bandwidth resources than what they need, and the broadband users do not have sufficient bandwidth.
To this scenario, game theoretic approaches based on the strategic form can be applied ideally as explained  in the following. The sensors will be individually allocated to orthogonal bandwidth resources, such as subcarriers and time slots,  whereas the broadband users join in these groups occupied by the sensors in a distributed manner. Since the number of the groups is determined by the number of the sensors, an action for one user jumping from   group $\mathcal{S}_i$ to group $\mathcal{S}_j$ will have no impact on the users in group $\mathcal{S}_k$, $k\neq i$ and $k\neq j$. Therefore, the proposed game is in the strategic form, for which those well-known approaches, including merge-and-split algorithms or  hedonic games, can be applied. Note that   this simplified game model will also enable the joint design of user grouping and resource allocation illustrated in Fig. \ref{system_model_pdnoma}, where resource allocation for different categories of users can be decoupled.

\section{Game Models for Resource Allocation in Code-Domain NOMA}

This section mainly discusses game-theoretic models to solve the resource allocation problems for CD-NOMA networks in the uplink and downlink scenarios.

\subsection{Uplink Noncooperative Contention Control Game}

In uplink CD-NOMA, mobile users can contend for one or more resources, which may pose great challenges for stringent latency requirements. In the conventional contention algorithm, when users want to transmit, they first go through the request and grant procedure in order to avoid the collisions. When two or more users try to transmit at the same time on the same channel, they are treated as non-orthogonal collision and the transmissions are considered unsuccessful. Specifically, if the spectrum is detected idle, the user will not access the and transmits immediately in that spectrum. The user, instead, will randomly choose a time slot from the set of contention free periods, and then, wait till the selected time slot is available for transmission. As the spectrum is shared among multiple mobile users, the collision will unavoidably occur if more than one users get access to the spectrum simultaneously. It is therefore convenient to formulate the radio resource allocation in a distributed way without the centralized control at the BS, while achieving the required performance in the network level.

To reduce congestion, decrease collisions, and guarantee fairness, as introduced earlier, an efficient distributed medium access protocol is necessary to share the radio resources efficiently. To this end, game theory is a proper tool used to design the distributed protocols for CD-NOMA networks. The contention based MAC can be formulated by a random access game, in which the multiple mobile users want to maximize their utilities individually by sharing a common channel. A selfish user typically waits for a smaller backoff interval to increase its opportunities of accessing the medium, which hence may reduce the channel access chance of those well behaved users. The difference with traditional carrier sense multiple access protocols is to also consider the selection of the proper frequency segment and the spreading sequence for users to get access, which can improve the connectivity possibility. The resource allocation problems in this scheme are related with contention window length adjustment, power control, and subcarrier allocation, etc.

\subsection{Downlink Matching Game model}

\subsubsection{Matching Theory Basics}
In economics, matching theory is a mathematical tool to describe the formation of mutually beneficial relationships over time \cite{7AM-1992}, which has been especially influential in labor economics used to describe the formation of new jobs and other human relationships such as marriages. Beyond its wide use in economics, matching theory has also been recently applied for analyzing wireless markets and studying the interaction of competitive agents in a network, such that the agents belong to two disjoint sets. It discusses the general cases that multiple agents on one side of the game are rivals to match with
multiple agents on the other side.

The matching problems can be in general defined below:
\begin{enumerate}
\item Bipartite matching with two-sided preferences: in this class, all the agents are split into two disjoint sets, in which the members rank a subset of the members of the other based on preference.
\item Bipartite matching with one-sided preferences: here, the participating agents are partitioned into two disjoint sets, in which only one set of the agents have preference over the other in a one directional manner.
\item Non-bipartite matching with preferences: all the agents form one single homogeneous set, in which each agent sets ranking to a subset of the others according to preferences.
\end{enumerate}
In addition to the way of classification above, the matchings can be also classified by the number of agents in each side, i.e. one-to-one matching, one-to-many matching, and many-to-many matching, etc.

\subsubsection{Matching Theory for User and Subcarrier Pairing}

\begin{figure}[!t]
\centering
\includegraphics[width=3.6in]{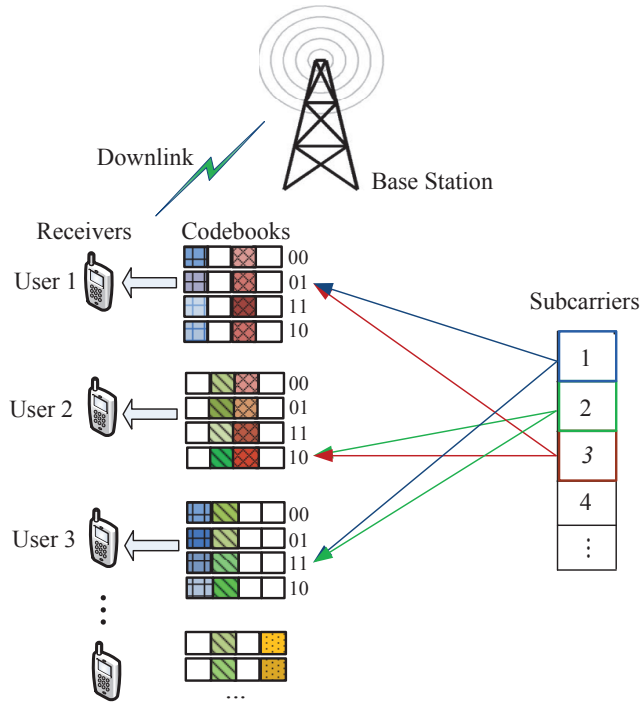}
\caption{System model of a CD-NOMA network.} \label{system_model_cdnoma}
\end{figure}

Consider a downlink single-cell CD-NOMA network£¬given in Fig.~\ref{system_model_cdnoma}.  A set of mobile users, denoted by ${\cal{N}} = \left\{ {1, \cdots ,N} \right\}$, communicate with a single BS, which divides the available bandwidth into a set of subcarriers, denoted by ${\cal{K}_{SC}} = \left\{ {1, \cdots ,K} \right\}$. There are $J$ codebooks assigned by the BS, and each user is characterized by a codebook that has $M$ sparse codewords with length $K$. At the transmitter, bit streams with length of $lo{g_2}M$  of each user are directly mapped into sparse codewords of the corresponding codebook, whose codewords can be generated by a multi-dimensional constellation. After mapping the data stream of each user to a sparse codeword, different CD-NOMA codewords are multiplexed over $K$ shared subcarriers for transmission. The objective of system design can be formulated to maximize the total sum-rate of the system through optimally allocating subcarrier to the users. This optimization problem can be treated as a many-to-many two-sided matching problem, and can be solved by utilizing the algorithms in matching theory~\cite{di-2016}.

To further illustrate the dynamic interactions among the users over shared subcarriers, i.e., how a specific codebook is designed for each user, in the formulated matching game, the set of users and the set of subcarrier can be considered as two disjoint sets of selfish and rational agents to maximize their own benefits, i.e. sum-rate. If subcarrier ${SC}_k$ is assigned to user $N_j$, then we say $N_j$ and ${SC}_k$ are matched with each other and form a \emph{matching pair}. The proposed algorithm can be described in three steps. First, the matching lists of each player are produced to record the preference. Step 2 implements the traditional deferred acceptance algorithm, in which each user proposes to its most preferred subcarrier which have not rejected it yet, and each subcarrier keeps its favorite offers and reject the others. The whole process stops when no user is willing to make new offers. Then we obtain an initial matching as input of Step 3. In Step 3, each matched user searches for another one to form a swap-blocking pair, and then exchange their matches and update the current matching. The process will continue iteratively until not any user can find new swap-blocking pairs. Finally, the final matching is determined~\cite{di-2016}.

\begin{figure}[!t]
\centering
\includegraphics[width=4.2in]{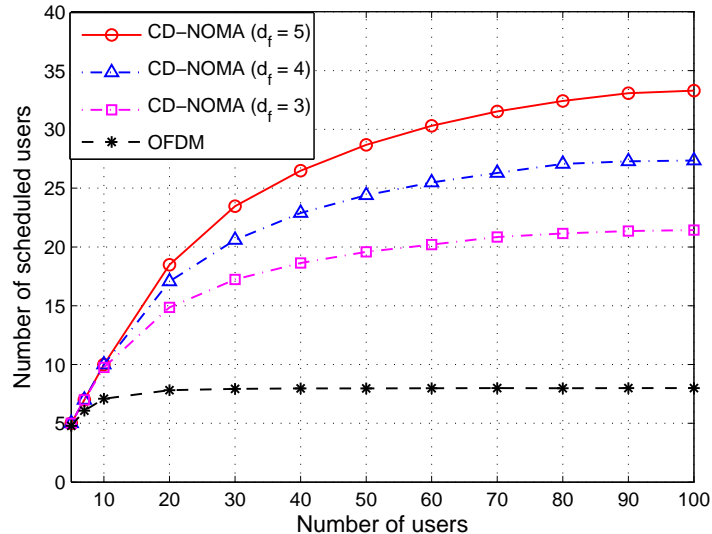}
\caption{Number of scheduled users v.s. number of the users $K = 8$.} \label{user_user_num}
\end{figure}
Fig.~\ref{user_user_num} shows the number of scheduled users vs. the number of users with $K = 8$ within one time slot, where $d_f$ represents the number of overloading users. It shows that the number of scheduled users grows with the number of users and is much larger than the number of subcarriers. In the OFDM scheme, at most $K$ users have access to the subcarriers, since one subcarrier can only be assigned to one user. When the overloading factor grows from 1 to 1.67, the number of scheduled users increases since more users can share the same subcarrier.

Table~\ref{tab:smmary_game} summarizes the game-theoretic models used for radio resource allocation of various NOMA schemes. Typically, non-cooperative game approaches such as power control result in suboptimal solutions but require smaller information exchange compared with many other cooperative game methods in terms of computational complexity.
%%%%%%%%%
\begin{table*}[h!]
\begin{center}
\caption{Summary of resource allocation game models for NOMA systems}\label{tab:smmary_game}
\begin{tabular}{|m{20mm}|m{65mm}|m{65mm}|}
\hline \bf{Applications} & \bf{Downlink Game Models} & \bf{Uplink Game Models}\\
\hline PD-NOMA & Coalition formation game: mobile users are categorized according to their QoS requirements, and distributed coalition formation algorithms based on the strategic form can then  be applied.  & Noncooperative power control game: the players are the mobiles, and a utility function is the negative of the individual transmit power of the mobile user transmission\\
\hline CD-NOMA & Matching game: the set of users and subcarrier can be considered as two disjoint sets of selfish and rational agents that aim to maximize their own benefits & Noncooperative contention control game: mobile users as players, pricing used to introduce cooperation, and contention window length adjustment for mobile users by iteratively updating algorithm\\
\hline
\end{tabular}
\end{center}
\end{table*}
%%%%%%%%%

%%%%%%%%%%%%%%%%%

\section{Conclusions and Future Directions}

In this article, we have discussed the design and resource allocation for NOMA systems to improve   network access efficiency. We have focused on PD-NOMA and CD-NOMA, two dominant  NOMA schemes. For PD-NOMA, coalitional game modes can be used in the downlink and non-cooperative power control games can be applied to the uplink to realize user and subcarrier grouping. While for CD-NOMA, matching games can be adopted in the downlink to pair the users,  and contention control games provide  an efficient tool to solve the uplink collision problems.

As an emerging technology, there are many open research problems in this field.
\begin{itemize}

\item Hybrid NOMA: NOMA encourages  multiple  users to share the same channels at the same time based on their  channel conditions. Thus, in the case in which  users's channel conditions are similar,  the performance gain of NOMA over OMA might be diminishing. Note also that the implementation of NOMA causes more complexity in comparison with OMA.  This motivates the design of hybrid NOMA where different MA  modes are used depending on the channel conditions. This process can be well modeled by evolutionary games: each mobile user is a player, and the strategies associated with each player are the algorithm which decides the proper mode the node should switch. A player may decide  to switch to a different strategy/operation mode if it can obtain a higher payoff. The change of strategy according to the payoff can be well modeled as the evolution process. The solution of the game is the evolutionary equilibrium, where, at the equilibrium, the players have no incentive to choose  different strategies.

\item MIMO NOMA: The basic idea of NOMA can be easily extended to the case in which all nodes are equipped with multiple antennas, which results in MIMO PD/CD-NOMA \cite{ZD-20161}. For downlink transmissions, the BS can use its multiple antennas either for beamforming to increase the receive SINR   or for spatial multiplexing to increase the throughput. While in the uplink, the objective of the NOMA design is to increase the spatial multiplexing gain using multiple antennas. In this case, the cross-layer optimization problem of MIMO applications becomes more complicated. Hence, how to effectively coordinate the use of resources at the space, time, frequency, and power domain is a challenging issue. Multi-level game models can be used to realize the joint resource allocation in different domains. For instance, a noncooperative game model can be applied for power allocation, while in the frequency domain,  an auction game can be used to assign frequency  resources.

\item Cooperative NOMA: relaying has been already deployed in many communications systems, such as LTE networks,  where multiple source nodes communicate with a destination with the help of one or multiple relays. By implementing NOMA at the relay, the relay can first receive and then forward superimposed signals simultaneously, which improves spectral efficiency~\cite{zhang-2016}. The communication process can be briefly described as follows: (1) the sources first transmit signals to both the NOMA relay and the destination at a given subcarrier, and (2) the NOMA relay then forwards the received signals to the destination with a proper transmit power at a given subcarrier. Hence, the resource allocation problems lie in a joint design for source-and-destination node grouping, subcarrier and power allocation, which can be modeled and solved by using the efficient auction games, where the relay is the auctioneer which provides power and spectrum resources, and the source nodes can be formulated as the bidders that have communication demands. The problem of power allocation for the relays to serve the sources can be also solved by using stakelberg game with one leader and multiple followers.

\item Cognitive NOMA: NOMA systems can be treated  as a special case  of cognitive radio networks, where users with poor channel conditions can be treated  as primary users. In order to suppress the interference temperature experienced at the primary users, sophisticated resource allocation algorithms can be designed by applying  stakelberg game, where NOMA users  can be categorized as either leaders or followers, according to their channel conditions or their QoS requirements. On the other hand, the concept of NOMA can be  also  used to improve the spectral efficiency of cognitive radio networks. For example, a reward policy regarding proper payment transfer should be applied to motivate    collaboration among primary and secondary users, which yields an ideal application for a contract game.

\end{itemize}

\end{document}